Marzo-abril 1925: crónica de un mes agitado

# ALBERT EINSTEIN VISITA LA ARGENTINA

por ALEJANDRO GANGUI y EDUARDO L. ORTIZ

El paso por nuestro país del autor de la Teoría de la Relatividad conmocionó a los círculos científicos y a la opinión pública. La singular personalidad y el genio de Einstein lo convirtieron en centro de agasajos durante su estadía en la Argentina. Del periplo por estas tierras damos cuenta en este artículo.

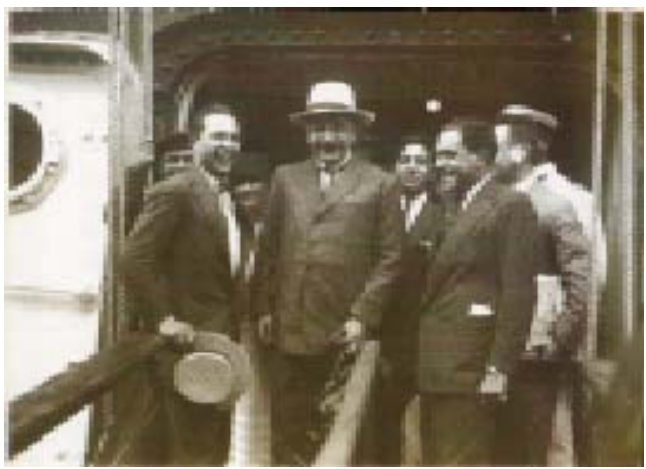

*La llegada de Einstein a la Argentina. "El eminente hombre de ciencia sonriendo a los fotógrafos en momentos de abandonar el barco que lo trajo a nuestra capital". (Actualidades de la semana: Revista* El Hogar, *marzo 1925).*

"Ciudad cómoda, pero aburrida. Gente cariñosa, ojos de gacela, con gracia, pero estereotipados. Lujo, superficialidad". Así es como Albert Einstein describió al Buenos Aires de 1925 y a su gente muy poco después de su llegada.

Durante su visita, el científico fue abrumado con interminables entrevistas, banquetes y honores: no tuvo demasiado tiempo para descansar. Llegó al puerto de Buenos Aires al alba del día 25 de marzo. Pero su periplo había comenzado casi tres semanas antes, cuando abandonó el puerto de Hamburgo a bordo del veloz y lujoso navío *Cap Polonio*, una muestra más de la potencia técnica de la Alemania de la primera post-guerra.

Su contacto con Argentina se había iniciado tres años antes cuando, por iniciativa del ingeniero Jorge Duclout, la Universidad de Buenos Aires (UBA) le cursó una invitación para dictar un ciclo de conferencias sobre su novísima y controvertida teoría de la relatividad general. Duclout, uno de los campeones de la teoría de la relatividad en Argentina, era un físico e ingeniero de origen francés radicado desde muchos años atrás en Buenos Aires y que, como Einstein, había estudiado en el Politécnico de Zurich.

Otras universidades y entidades argentinas se habían adherido a esta invitación. Entre ellas se destaca la Asociación Hebraica Argentina (hoy Sociedad), recientemente creada por una primera generación de intelectuales argentinos de origen judío que habían elegido la figura de Einstein como emblema de sus aspiraciones sociales e intelectuales.

Luego de hacer una escala breve en Río de Janeiro, Einstein _que viajaba solo_ realizó el trayecto Montevideo - Buenos Aires acompañado por miembros de diversas comitivas de bienvenida que habían ido a esperarlo a la otra orilla del Plata. Los integrantes eran científicos, personalidades académicas y miembros de la comunidad judía de Argentina. Entre ellos se encontraba el secretario de la Universidad de Buenos Aires, Mauricio Nirenstein, que era también miembro de la Asociación Hebraica.

A su llegada al puerto de Buenos Aires Einstein recibió una primera muestra del impacto público de su visita. Una masa de periodistas, cámaras y filmadoras lo aguardaban en la dársena de desembarco. Apresurados por esquivarlos, los acompañantes de Einstein intentaron evadirse con él en un automóvil particular, pero fueron firmemente detenidos por los periodistas que usaron el propio equipaje de los viajeros para bloquear el camino. No los dejaron partir hasta haber fotografiado y filmado a Einstein a su gusto en el interior del vehículo.

Dice Einstein en su diario íntimo: "A las 8:30 estábamos en tierra firme. Nirenstein me prestó ayuda". Aclara luego que se sentía "medio muerto" luego de tanto viaje y ajetreos entre la multitud. Finalmente llegó a la residencia de Bruno Wassermann, un rico comerciante de origen judío-alemán, situada en la parte más elegante de Belgrano; allí se alojó durante su estadía en Buenos Aires. Finalmente Einstein pudo tomarse un brevísimo descanso y escribió en su diario: "Tranquilidad por fin, estoy totalmente deshecho". Una vez que Einstein llegó a Buenos Aires, Nirenstein compartió la secretaría universitaria con la función de secretario del ilustre visitante, o quizás más bien de severo preceptor.

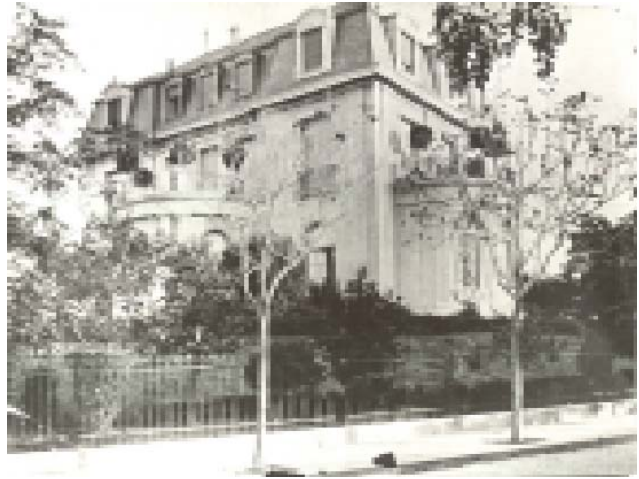

*La residencia de la familia Wassermann, donde Einstein se alojó durante un mes.*

Bienvenido señor Einstein

Las inevitables visitas protocolares no se hicieron esperar. Dejando de lado a los periodistas, la lista de los intelectuales que lo visitaron fue encabezada por un poeta, Leopoldo Lugones, que era uno de los pocos argentinos, y quizás el único intelectual argentino, a quien Einstein había conocido antes de su viaje. Einstein había compartido con Lugones, Marie Curie, Henri Bergson y otras figuras eminentes, la mesa de discusiones de uno de los foros intelectuales más importantes de esa época: el Comité Internacional de Cooperación Intelectual de la Liga de las Naciones; Lugones representaba allí a Argentina. Este era un embrión pretérito de UNESCO.

Einstein ha declarado que conocía el texto "posmoderno" de Lugones: *El tamaño del espacio: Ensayo de Psicología Matemática* (El Ateneo, Buenos Aires, 1921), desde el año de su publicación[1]. El mérito principal de este libro no está en la física, que Lugones claramente no dominaba ni presumía dominar, sino en haber señalado al Buenos Aires intelectual de su tiempo que no sería inoportuno prestar atención a las preocupaciones de Einstein por comprender mejor los conceptos de espacio, tiempo, materia y energía.

Siguió luego una larga lista de personalidades con las que se entrevistó: el embajador alemán, autoridades de la UBA, algunos científicos de renombre, representantes de organizaciones judías y no judías (como la Asociación Cultural Argentino-Germana) que habían colaborado en la financiación de su circuito en el país y que, con éxito variable, habían intentado extenderlo a otros lugares de Sudamérica.

El diario *La Prensa* describió a Einstein utilizando un estereotipo desarrollado ya por el periodismo de otros países. Encontró que era "un hombre bondadoso, afable y simpático", famoso a causa de ser el "autor de una teoría científica que ha llamado la atención del mundo". A estos atributos se agregarían luego su humildad, la pobreza de su vestimenta y el rol importante que en su presencia física jugaba su impresionante y desplegada cabellera.

En declaraciones al mismo matutino, Einstein agradeció primero "la elogiosa crónica que sobre mi persona publicó ayer el gran diario argentino", para manifestar enseguida que era "enemigo del exhibicionismo" y que "me consideraría muy satisfecho si no se me abrumara tanto con el sinnúmero de entrevistas que se me solicitan" (*La Prensa*, 26/03/1925).

No todo es ciencia

La primera contribución de Einstein, escrita especialmente para una publicación Argentina, no fue científica. En el día previo al de su llegada _en la edición del 24 de marzo (página 14)_ *La Prensa* publicó un artículo de Einstein titulado "Pan-Europa"[2]. En ese trabajo Einstein hizo una crítica al nacionalismo, defendió el renacimiento de la comunidad europea y apoyó la postura de quienes, como él, se esforzaban por lograr una integración de Europa, por lo menos en el mundo de la cultura. En el artículo resumió las preocupaciones contemporáneas del mundo intelectual de Berlín y de Europa. Esta fue su carta de presentación al público culto de Argentina.

Al comienzo de su visita Einstein se mostró dispuesto a responder a preguntas de todo tipo. Parecía preparado para discutir tanto sus impresiones de viaje o de Buenos Aires, como temas espinosos y delicados. Por ejemplo, las consecuencias de la Guerra Mundial sobre la unidad de Europa; las perspectivas de una paz duradera; los problemas nuevos que presentaba el socialismo; los problemas de los judíos en Alemania y fuera de ella, y un sinnúmero de temas de similar interés. Su relación con la prensa local fue abierta y cordial; accedió incluso a una entrevista más larga que lo normal con un reportero que se presentó pobremente vestido y necesitado de afianzar su posición en el periódico: Einstein le regaló un retrato suyo autografiado.

Sin embargo, esa postura amplia y generosa de Einstein no se prolongó más allá de los primeros días. Rápidamente sus contactos con la prensa argentina habrían de tornarse más distantes y medidos. Ninguno de los temas espinosos volvería a ser tocado a lo largo de su visita. Esta deficiencia sugiere que inicialmente hubo de su parte la propuesta de una agenda amplia, pero que ella, en una gran medida, quedó incumplida en Argentina.

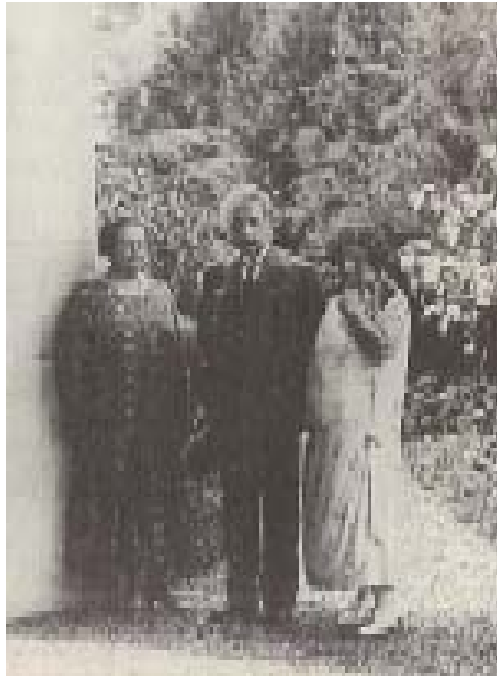

*Einstein junto a la señora Wassermann (a la derecha del científico) y la escritora Elsa Jerusalem, en los jardines de la residencia donde se hospedó.*

Una agenda abultada

La mañana del jueves 26 de marzo volvió a recibir a los periodistas y fotógrafos en la residencia de los Wassermann. Al mediodía, lo llevaron a conocer dos caras diferentes de la ciudad: los cuidados bosques de Palermo _nuestro *Bois de Boulogne*_ y el Mercado de Abasto Proveedor, que Einstein cruzó a pie con su comitiva. Esta última visita evocaba la imagen del tango argentino que comenzaba ya a atraer la atención de Europa. A la tarde asistió a una recepción formal en el salón de actos del Colegio Nacional de Buenos Aires. Es allí donde Einstein desarrollaría su ciclo de conferencias. La primera de ellas, pronunciada en francés ante un público amplio y numeroso, fue breve y de carácter general. Su primera lección estrictamente científica en Buenos Aires (de un total de ocho; las autoridades de la UBA habían convenido con Einstein un total de doce conferencias a ser dictadas en diferentes centros académicos) se realizó el sábado siguiente.

El viernes por la noche, fue invitado a una recepción en la residencia de un miembro influyente y adinerado de la comunidad judía local; allí se encontró con personas que habían apoyado financieramente su viaje a Argentina.

El sábado 28 comenzó su serie de conferencias. La sala estaba colmada, escribe Einstein en su diario, agregando: "La juventud es siempre agradable y se interesa por las cosas". A esta conferencia, además de las autoridades universitarias del más alto rango, asistieron dos ministros del gabinete, los titulares de las carteras de Educación y de Relaciones Exteriores y varios embajadores extranjeros. Al día siguiente, el científico logró "permanecer solo en [su] habitación durante la mañana" y disfrutar de una jornada de "feliz tranquilidad". Por la tarde, hizo una caminata en compañía de Bruno Wassermann.

Pero llegado el lunes el protocolo continuó con paso firme: al mediodía visitó los modernísimos talleres del diario *La Prensa*; luego dictó su segunda conferencia científica sobre relatividad, con abundante tiempo para discusiones. Varios científicos locales intentaron mostrarse capaces de hacer preguntas, no siempre con un éxito rotundo. Al día siguiente, martes 31, por la mañana visitó la redacción del diario judío *Das Volk*, e hizo observaciones picantes acerca de sus "paisanos". Visitó también escuelas, hospitales y orfanatos sostenidos por esa comunidad. Por la tarde viajó al Tigre invitado por unos amigos suyos suizo-alemanes.

El miércoles 1º de abril Einstein realizó un vuelo corto sobre la ciudad de Buenos Aires a bordo de un junker de la marina alemana que había llegado a Buenos Aires en vuelo de cortesía; lo acompañó la señora Wassermann. Este era su primer vuelo en avión; Einstein comentó luego cuánto lo impresionó esa experiencia, "particularmente el despegue" de la aeronave. Por la tarde fue recibido por el presidente de la república, Marcelo T. de Alvear, y por algunos ministros. Visitó luego el Museo Etnológico y, poco después, se encontraba dictando su tercera conferencia sobre la Teoría de la Relatividad. La jornada terminó con un paseo a pie por Florida, del brazo de Lugones y seguido de estudiantes y curiosos. La caminata concluyó en casa del poeta, donde cenaron con la esposa de Lugones. Fue la única oportunidad de Einstein de compartir una cena íntima en un hogar argentino.

Al día siguiente viajó en tren a La Plata, donde había sido invitado a inaugurar el año académico de 1925; asistió también a una reunión científica en su honor, donde participaron el físico alemán Ricardo Gans, y algunos de sus alumnos. Logró hacerse de tiempo para visitar el "muy interesante Museo de Historia Natural". El viernes 3 de abril, fue invitado por las autoridades de la UBA a almorzar en el Jockey Club, y luego dictó otra conferencia más de su ciclo para la universidad. Al día siguiente le tocó el turno a la Facultad de Filosofía y Letras de la UBA, donde repitió la función que dos días antes había cumplido en La Plata. Sin embargo esta vez, a instancias del decano Coriolano Alberini, Einstein tomó parte activa ofreciendo un coloquio breve sobre "Las consecuencias de la teoría de la relatividad en los conceptos de espacio y tiempo". Con satisfacción, Alberini interpretó su conferencia como un golpe contra el positivismo.[3]

El domingo 5, con los Wassermann, viajó en automóvil a la residencia de vacaciones que esa familia tenía en Lavallol. Fue un día corto pero reposado. El lunes 6 de abril continuó su contacto con científicos locales: en compañía del joven fisiólogo Bernardo A. Houssay (que 22 años más tarde sería galardonado con el Premio Nobel de Medicina), visitó el laboratorio del "oftalmólogo y bolsista" de origen francés doctor Eugenio Pablo Fortin, que estaba haciendo interesantes experimentos sobre la percepción de sensaciones luminosas. Por la tarde dictó una nueva conferencia en su ciclo para la UBA; terminó el día participando en una reunión pública organizada por la comunidad judío-argentina para festejar la muy reciente inauguración de la Universidad Hebrea de Jerusalén, y dinamizar la recolección de fondos para esa institución. A su muerte Einstein legó a esa universidad los originales de sus documentos escritos, antes depositados en el Archivo de la Universidad de Princeton.

La visita a la clínica universitaria en compañía del doctor José Arce, rector de la UBA, quedó para el día siguiente; Arce y el establecimiento le dejaron excelente impresión. El miércoles 8 de abril, Einstein decidió adelantar el receso de Semana Santa de ese año y volver, con los Wassermann, a la estancia de Lavallol. Fuera del ruido de la ciudad, encontró allí "hermoso clima, maravillosa quietud", como dejó anotado en su diario; escribió también que había tenido "una idea sobre una nueva teoría sobre la conexión entre la gravitación y el electromagnetismo".

Sin embargo, su viaje a la Argentina no dejaba mucho tiempo para pensar en nuevas teorías. El día sábado 11 de abril, junto a varias personalidades locales _el ingeniero Enrique Butty, los físicos Ramón Loyarte y Teófilo Isnardi, y los decanos de Ingeniería, Luis A. Huergo, y de Filosofía Alberini_, Einstein abordó vagones especiales del tren nocturno a Córdoba. La prensa local anunció por anticipado que "[Einstein] será recibido [en Córdoba] como embajador espiritual de la nueva Alemania". Inmediatamente después de su llegada, las autoridades universitarias y provinciales lo invitaron a dar un paseo por las sierras, visitando luego el Lago San Roque para terminar almorzando en el Edén Hotel de La Falda; regresaron por el camino de Alta Gracia. De nuevo en la ciudad, pudo admirar la Catedral y los restos de lo que en su diario calificó como una antigua cultura. El lunes 13 de abril disertó durante una media hora acerca del desarrollo de la Teoría de la Relatividad: la teoría restringida, la teoría general y los esfuerzos que contemporáneamente se hacían por poner la gravitación y el electromagnetismo dentro de un mismo esquema teórico.

Su amigo el doctor Georg Friedrich Nicolai, pacifista como Einstein y que hasta tres años antes había sido profesor de fisiología en la Universidad de Berlín, enseñaba ahora en Córdoba. Aunque ambos amigos se encontraron, es poco lo que se puede decir de esta entrevista. Curiosamente su diario íntimo no hace referencia alguna a ese encuentro. ¿Era su diario realmente íntimo?

A su llegada Einstein había expresado interés por visitar las colonias judías de Entre Ríos, si el tiempo se lo permitía; ese deseo no pudo ser cumplido. En cambio, para su regreso de Córdoba a Buenos Aires, eligió hacer un viaje diurno, partiendo de Córdoba a las 6:45 de la mañana del día martes 14 para, por lo menos, poder ver el sur de esa provincia y parte de Santa Fe.

**RECUADRO: Un sabio, una autoridad**

Einstein fue mundialmente conocido sobre todo por sus teorías de la relatividad: la relatividad restringida (1905) y la relatividad general (1915). En esta última, Einstein ofreció un nuevo marco teórico que reemplazó y mejoró la teoría de la gravitación universal de Newton[1]. Aunque inicialmente muy pocos entendían de qué se trataba esta nueva visión del universo -más propia de la ciencia ficción que de la realidad cotidiana-, la revolución del espacio-tiempo einsteniano caló hondo en la sociedad. Cuando Einstein llegó a la Argentina, su fama y su nombre eran sinónimos de «la relatividad». En su edición del 28 de marzo de 1925, la revista *Caras y Caretas* lo describió con las elogiosas palabras: "El famoso sabio alemán es hoy una figura mundial. Su célebre teoría de la relatividad revolucionó las bases mismas de las ciencias...".

Lentamente, y muy a su pesar, Einstein se convirtió en una leyenda viviente, en un héroe popular y en una autoridad en todo tema donde se buscara una opinión de peso. Justo le sucedía esto a él, que desde pequeño en sus estudios había mostrado rebeldía ante toda forma de autoridad, especialmente en lo que se refería a esos rígidos y repetitivos métodos alemanes de formación escolar. Con cierta cuota de ironía, Einstein diría años más tarde: "Para castigarme por mi desprecio a la autoridad, el destino me convirtió a mí mismo en una autoridad".

1. Gangui Alejandro, *El Big Bang: la génesis de nuestra cosmología actual*. Buenos Aires, Eudeba, 2005. Considera los problemas que surgen con la Teoría de la Gravitación Universal de Newton hacia fines del siglo XIX, y el surgimiento de las Teorías de la Relatividad de Einstein, y sus consecuencias cosmológicas.

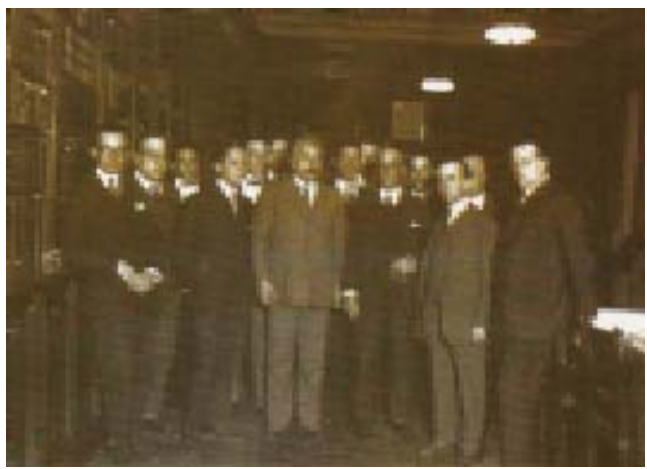

*Einstein en la biblioteca de* La Prensa, *rodeado por las autoridades del diario, durante su visita del mediodía del lunes 30 de marzo.*

Einstein, académico argentino

Nuevamente en Buenos Aires, en la mañana del 16 de abril Einstein se reunió con dirigentes de la Federación Sionista Argentina. A la tarde asistió a una sesión especial de la Academia Nacional de Ciencias Exactas, Físicas y Naturales en su honor, en la que el presidente, el naturalista y escritor doctor Eduardo L. Holmberg, le entregó su diploma de Académico Honorario. De acuerdo con la carta de invitación de Holmberg, a continuación se pasó a escuchar "su ilustrada palabra en respuestas o preguntas que, sobre la Teoría de la Relatividad y problemas afines, le formularán algunos miembros de la Academia y otras personas invitadas especialmente". Los *Anales de la Academia* y otros textos históricos coinciden en destacar la intervención del joven físico uruguayo Enrique Loedel Palumbo, que fue luego maestro de importantes científicos de la actualidad. Su pregunta, relacionada con la existencia de una solución para un sistema de ecuaciones que describe el campo gravitacional de una masa puntual, y que Einstein encontró de interés, dio origen a una publicación en la revista alemana *Physikalische Zeitschrift* el año siguiente[4]. La nota sobre esta reunión en el diario de Einstein no deja demasiado bien parados a sus interlocutores.

El 17 de abril, por la tarde, Einstein dictó su penúltima conferencia en la UBA. Esa noche fue agasajado por el embajador alemán, en una reunión que comentaría en su diario señalando que casi "no había alemanes". Entre los invitados argentinos se encontraban el doctor José Ingenieros, el músico Carlos López Buchardo, el escritor Calixto Oyuela, el escultor Rogelio Yrurtia, el ingeniero Nicolás Besio Moreno, y otros; del lado argentino-alemán: miembros de la embajada, Bruno Wassermann y Ricardo Seeber, que entonces era presidente de la Asociación Cultural Argentino-Germana.

Recordemos que Einstein había renunciado a su ciudadanía alemana antes de cumplir los 16 años, la edad de registro militar, y que más tarde, en 1901, adquirió la nacionalidad suiza. Al ser incorporado a la Academia Prusiana de Ciencias implícitamente se le había reconocido nacionalidad alemana, aunque él nunca abandonó la adoptada. El desaire que Einstein comenta en su diario tiene antecedentes concretos.

Al desatarse la Primera Guerra Mundial unos cien importantes intelectuales alemanes firmaron un manifiesto dando su apoyo firme a la guerra; Einstein, profesor en la Universidad de Berlín, firmó un contra-manifiesto denunciando la guerra. El documento de Einstein sólo tiene el apoyo suyo y el de otros tres nombres. Nicolai, el profesor recluido en Córdoba, era el primero de esos tres. A pesar de haber enfrentado numerosas dificultades, Einstein nunca dejó de reafirmar su fe pacifista. En su diario íntimo volcó su disgusto comentando: "gente rara estos alemanes. Para ellos soy una flor maloliente, y sin embargo, una y otra vez, me ponen en su ojal".

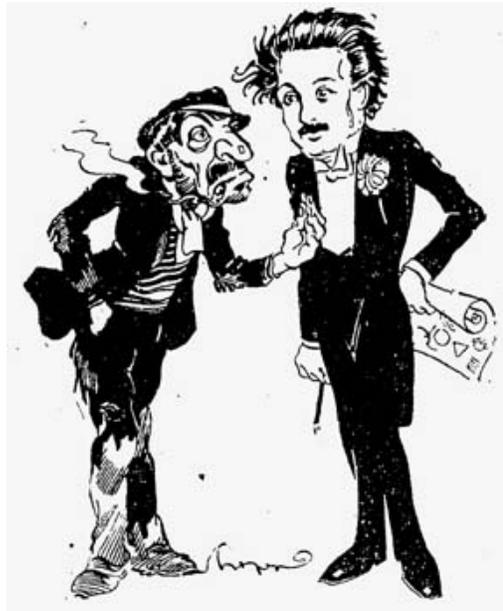

*Ilustración de la nota de* Last Reason *(seudónimo del periodista Máximo Teodoro Sáenz) publicada en el diario* Crítica *en marzo 1925, previamente a la llegada de Einstein al país. Las crónicas de Sáenz aparecían siempre en la última página de* La Razón*, de ahí su* nom de plume.

El sábado 18, por la tarde, Einstein ofreció una conferencia privada sobre su teoría en casa de sus anfitriones. Hacia la noche asistió, como invitado de honor, a una recepción en el cine-teatro Capitol donde expuso "Algunas reflexiones sobre la situación de los judíos". A continuación la Asociación Hebraica Argentina ofreció una recepción en sus salones y entregó a Einstein el diploma de socio honorario. El domingo Einstein viajó nuevamente a Lavallol; por la noche se reunió con amigos alemanes residentes en Buenos Aires y luego asistió a una recepción en su honor ofrecida por dirigentes judeo-argentinos en el Savoy Hotel.

Aún le quedaba por dar una conferencia: el lunes 20 "cumplió con su contrato" anotando en su diario que había cumplido con su "última sesión científica con una audiencia entusiasta". También se hizo tiempo ese día para visitar al ingeniero Duclout, que se encontraba enfermo. Como si lo ya enumerado fuese poco, el 21 de abril Einstein visitó el Hospital Israelita y otras organizaciones de caridad de la colectividad. Al mediodía el Consejo Directivo de la Facultad de Ciencias Exactas lo invitó a almorzar en el Yacht Club Argentino, en el Tigre.

Logró también hacerse tiempo para componer algunos poemas breves que insertó como dedicatoria en fotografías suyas que luego obsequió a algunas de las personas con las que había hecho relación durante su viaje: la escritora Elsa Jerusalem, esposa del profesor de embriología Víctor Widakowich, contratado por la Universidad de La Plata, a quien conoció en el *Cap Polonio* y con quien siguió manteniendo amistad; la señora de Wassermann y el profesor Nirenstein entre otros. En sus dedicatorias, en alguna medida, define los roles que ellos jugaron en su visita y les expresa su simpatía y agradecimiento. La dedicatoria a Nirenstein sugiere que éste jugó un papel importante en "moderar" a Einstein durante su visita a Argentina, "para que posiblemente nadie se sintiera ofendido" como dice la dedicatoria.

El miércoles 22 fue invitado a un almuerzo de despedida organizado por la cúpula científica y política de Argentina en el Jockey Club, donde participaron rectores, decanos y ministros. El cierre de la jornada fue menos formal y algo más divertido: esa noche Einstein asistió a una fiesta organizada por el Centro de Estudiantes de Ingeniería en la Asociación Cristiana de Jóvenes, que el homenajeado describió como "estudiantes, guitarras y canto". Entusiasmado por la cordialidad de la audiencia, Einstein aceptó tocar el violín; "con la velocidad de la luz", nos dice la crónica, un estudiante disparó a su casa a buscar el instrumento.

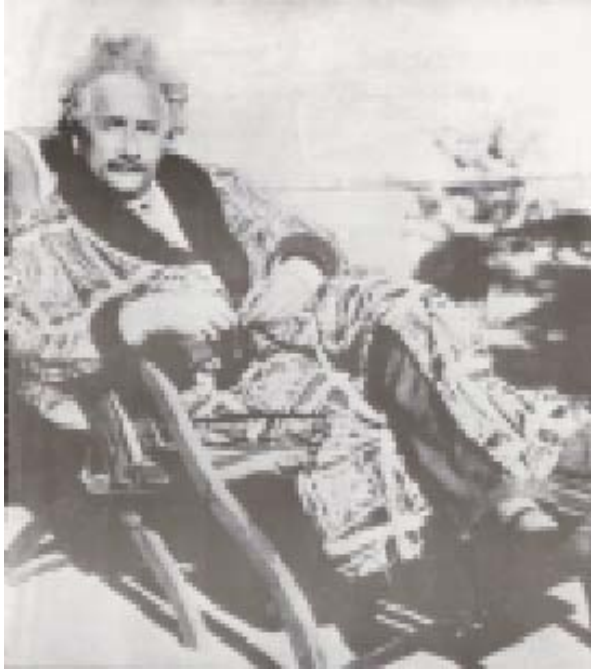

*Albert Einstein de entrecasa reposando al sol.
(Archivo General de la Nación).*

**RECUADRO: Impresiones periodísticas**

Los primeros reportajes de la prensa local hechos a Einstein tienen el mayor interés, tanto por los temas discutidos, como por las primeras visiones que los periodistas nos ofrecen del científico en la escena argentina. Una de esas entrevistas la hizo un periodista del diario *Crítica* en los primeros días de residencia de Einstein en casa de los Wassermann.

La residencia de Bruno Wassermann, que estaba en consonancia con su crecida fortuna, no pareció a los periodistas el escenario más adecuado para un Einstein a quien veían, a la vez, como un hombre humilde y modesto en sus costumbres, y como un revolucionario en su pensamiento. Quizás también Einstein puede haberse sentido algo fuera de ambiente en esa mansión; en Montevideo eligió vivir en una residencia considerablemente más modesta, que le ofreció un pariente de amigos suyos.

El reportero de *Crítica* cuenta que al traspasar la puerta de la residencia se encontró "en un salón magnífico que ocupa gran parte de la planta baja del palacio y a poco rato vimos aparecer la figura característica del sabio, vestido con un traje cruzado abotonado hasta el cuello" y con "las sandalias que tanto chocaron a los viajeros de primera [clase] del *Cap Polonio*."

Pasada esa primera impresión, sólo dictada por su calzado, Einstein, "con un saludo afable y franco se aproximó a nosotros buscando con la mirada un sitio adecuado para poder platicar con calma. Al examinar con la mirada el gran salón se dibujó en su rostro una sonrisa picaresca, como diciendo que esto sólo sirve para grandes fiestas, pero no para coloquios quietos e íntimos. Al fin nos invitó a seguirlo a una puerta lateral y nos encontramos de pronto en un pequeño escritorio amueblado adecuadamente para poder recogerse y reflexionar con libertad".

Una de las primeras preguntas fue acerca de qué país consideraba Einstein que prestaba en ese momento una mayor atención a las ciencias. Sin vacilar, y siempre según el periodista, Einstein respondió que ese país era Alemania, donde desde la enseñanza secundaria se inculcaba a los alumnos el hábito de pensar con profundidad acerca de los conceptos de la ciencia pura. Esta respuesta resulta sorprendente, particularmente si se recuerda que Einstein dejó la educación alemana y su afamada escuela secundaria, en favor de la de Suiza.

En esos primeros días de residencia en Argentina, la escritora Elsa Jerusalem, "que no se separa de él", debía actuar como intérprete entre el sabio y un reportero del diario *Die Presse*; sin embargo, la comunicación fue más directa: la entrevista se condujo en *yiddish* del lado del periodista, y en alemán del de Einstein. Enseguida se entabló un interesante diálogo sobre el futuro de la lengua *yiddish*. Einstein creía que el hebreo, que iba a ser empleado en la enseñanza en la nueva Universidad de Jerusalén, era la lengua que tenía un mejor futuro. El periodista, que representaba a un diario porteño que se publicaba en el primero de esos idiomas, defendió su territorio e insistió que esa "era la lengua madre de millones de judíos que, en ella, gozan de la ciencia y del arte". Jugando con las palabras, Einstein replicó: "¿lengua madre? ¿Y por qué no [usar] la lengua abuela?". Ante la sugestión de que se estaba matando a esa lengua madre, Einstein respondió: "¿Y qué hacemos cuando se muere la abuela?". Ambos rieron y pasaron a ocuparse de otros temas.

Sin embargo, no todas fueron sonrisas en ese importante viaje. En el diario íntimo de Einstein hay también referencias duras, tanto acerca de los pasajeros argentinos de la primera clase del *Cap Polonio* durante el viaje a Buenos Aires, como de algunos de sus colegas locales _particularmente quienes le formularon preguntas científicas en la Academia de Ciencias_ y, también, acerca de miembros de la comunidad judía en Buenos Aires, que en ocasiones colmaron su paciencia.

---

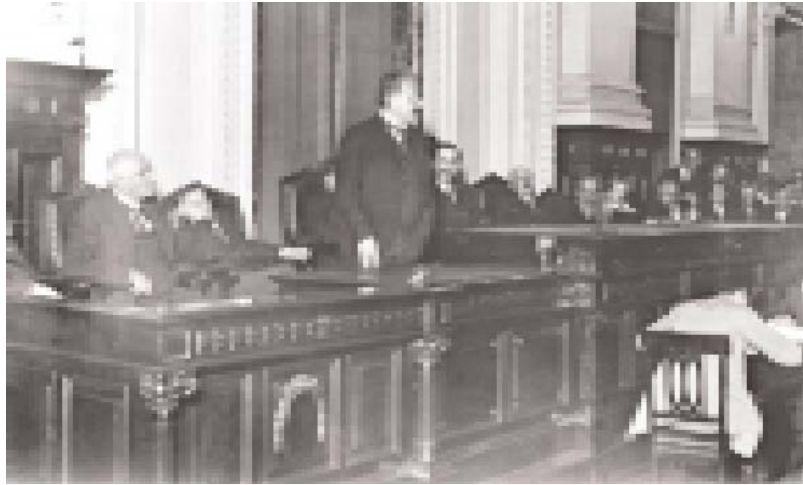

*Einstein durante la recepción que se le ofreciera el 26 de marzo en el salón de actos del Colegio Nacional de Buenos Aires. Dos estudiantes de ciencias (hacia la derecha de la imagen) toman nota de todas sus palabras.*

**RECUADRO: Aniversario de un "año milagroso"**

El año 2005 fue declarado el Año Mundial de la Física y conmemora la primera centuria de una serie de trabajos científicos legendarios de Albert Einstein (1879-1955). Contando con sólo 26 años, Einstein envió a publicar _en algo más de 6 meses_ cinco *papers* (uno de ellos, con el material de su tesis doctoral) que conmovieron los cimientos de la física de su tiempo y dieron origen a nuevas líneas de investigación con resultados entonces insospechados.

Los temas tratados explicaban observaciones pioneras en la naciente teoría cuántica y en la interacción de la luz con la materia, amalgamaban el estudio de las ondas electromagnéticas con la mecánica clásica galileana _corrigiendo a esta última_ y explicaban el movimiento errático de pequeñas partículas de polen inmersas en un líquido _el llamado "movimiento browniano"_ que permitiría poner en evidencia firme la existencia de los átomos como constituyentes básicos de la materia.

El 18 de marzo de 1905, las oficinas de Berlín de la revista *Annalen der Physik* reciben el primer trabajo de Einstein de ese año. Se titula «Acerca de un punto de vista heurístico referido a la generación y transmutación de la luz» y es publicado el 9 de junio siguiente[1]. En mayo de ese año, en una carta que envía a su amigo Conrad Habicht, Einstein hace mención de éste y de los otros trabajos científicos que luego se harían célebres: "Querido Habicht, [..] por qué no me ha enviado aún su tesis? No sabe usted que yo soy una de la persona y media en total que la leerían con interés y placer [..]? A cambio, le prometo cuatro de mis artículos. Y hasta podría enviarle el primero muy pronto, pues estoy por recibir los ejemplares gratuitos. Este trabajo trata sobre la radiación y las propiedades energéticas de la luz y es muy revolucionario…"[2]. Será por este trabajo «revolucionario», y no por la teoría de la relatividad, que a Einstein se le otorgará el Premio Nobel de Física del año 1921.

1. Einstein A., 1905, "Über einen die Erzeugung und Verwandlung des Lichtes betreffenden heuristischen Gesichtspunkt", *Annalen der Physik*, Leipzig, 17, pp.132-148. El texto en idioma original y su traducción al español puede ser consultado en el sitio de Internet: www.universoeinstein.com.ar

2. Einstein A., *The Collected Papers of Albert Einstein, Volume 5: The Swiss Years: Correspondence*, 1902-1914. Edited by M. J. Klein, A. J. Kox, and R. Schulmann, Princeton University Press, 1993. Puede sorprender que Einstein, de 26 años, tratase de usted a un amigo íntimo como lo era Habicht (y al mismo tiempo intercambiase algunas injurias pintorescas que no hemos traducido..). Esa era la regla de formalidad en la Suiza de principios del siglo XX, incluso entre jóvenes amigos.

La recta final

Es de esperar que al día siguiente Einstein se hubiese levantado más tarde que de costumbre, aunque, quizás por modestia, este dato no consta en su diario. Al mediodía los Wassermann ofrecieron en su casa un almuerzo de despedida, al que invitaron a los físicos Loyarte e Isnardi. El resto de ese día _el de su partida_ el célebre científico lo ocupó en empacar su reducido equipaje (aumentado con algunos regalos) y en recibir a algunas amistades a las que obsequió las mencionadas fotografías autografiadas. En la noche del 23 de abril Einstein dejó para siempre la ciudad de Buenos Aires y se embarcó hacia Montevideo, a donde llegó _según reportan las crónicas de la época_ con aspecto de cansado y sintiéndose algo enfermo. Allí ofrecería tres conferencias en la Facultad de Ingeniería, y sería invitado de honor en varias recepciones locales. Nuevamente respiró un aire de libertad del que parece no haber tenido suficiente en Buenos Aires, donde puede haber estado más estrictamente controlado. Por esta razón, o por otras, justo es decir que la impresión que su diario refleja de nuestros hermanos rioplatenses fue algo mejor que la que le dejaron los porteños. Luego de una semana, partió rumbo al Brasil en un buque al que describió como "muy sucio y pequeño, pero con una tripulación agradable". Luego de una estadía de una semana en Río emprendió su viaje de regreso a Europa, el 12 de mayo de 1925.

A su llegada a Alemania, fue entrevistado por su amigo Otto Buek, que además de ser corresponsal en Berlín de *La Nación*, era el segundo de los otros tres firmantes del Contra-Manifiesto pacifista de 1914.

Einstein pronosticó "un gran porvenir económico y cultural" para Argentina, manifestando asimismo que conserva "los mejores respetos de su hermoso viaje a la América del Sur" (*La Nación*, 5/06/1925). Su diario íntimo, sin embargo, no fue siempre fiel a sus declaraciones, o al reporte público de las mismas. Su diario en Sudamérica se cierra con una frase elocuente: "Al fin libre, pero más muerto que vivo". Setenta y cinco años más tarde, la revista *TIME* lo designó "personaje del siglo". Entre los viajeros científicos llegados a Argentina durante el siglo veinte, sin duda, Einstein ocupa esa misma posición. ☺

NOTAS

1. Asúa Miguel de y Hurtado de Mendoza Diego, "La conferencia de Leopoldo Lugones sobre el espacio y la teoría de la relatividad (1920)", *Ciencia Hoy*, Buenos Aires, Vol. 15, Nº 85, 2005, p. 62.

2. Einstein Albert, "Pan-Europa", *La Prensa*, Buenos Aires, 24 de marzo, p. 14. Reproducido en *Revista de Filosofía*, Buenos Aires, 1925, Vol. 11, pp. 468-470.

3. Alberini Coriolano, "Einstein en la Facultad de Filosofía y Letras", *La Nación (Suplemento)*, Buenos Aires, 12 de abril 1925. También puede verse Hurtado de Mendoza Diego, "Las teorías de la relatividad y la filosofía en la Argentina (1915-1925)", en Marcelo Montserrat (comp.), *La ciencia en la Argentina de entre siglos*, Buenos Aires, Manantial, 2000, pp. 35-51.

4. Loedel Palumbo Enrique, "Forma de la superficie espacio-tiempo de dos dimensiones de un campo gravitacional proveniente de una masa puntiforme", *Contribución al estudio de las ciencias físicas y matemáticas*, La Plata, 1926. Serie Matemático-Física, IV, pp. 81-87. (Este es el trabajo que Loedel Palumbo publicó en alemán como "Die Form der Raum-Zeit-Oberfläche eines Gravitationsfelde, das von einer punktförmigen Masse herrürht", *Physikalische Zeitschrift*, 27, 645-648).

REFERENCIAS SOBRE LA VISITA DE EINSTEIN A LA ARGENTINA

Agulla (h), Juan Carlos, "Einstein en la Argentina", *Todo es Historia*, Buenos Aires, Nº 247, enero de 1988, pp. 38-49.

Mariscotti Mario, "La visita de Einstein a la Argentina", *Revista de Enseñanza de la Física*, Buenos Aires, Nº 9 (1), 1996, pp. 57-66.

Nirenstein Mauricio, "Einstein en Buenos Aires", *Verbum, Revista del Centro de Estudiantes de Filosofía y Letras*, Buenos Aires, Nº 18, 1925, p. 168.

Ortiz Eduardo L., "A convergence of interests: Einstein's visit to Argentina in 1925", *Ibero-Amerikanisches Archiv*, Berlín, Nº 21 (1-2), 1995, pp. 67-126. Monografía basada en documentos de archivos nacionales y extranjeros. Considera la gestación y el desarrollo de la visita a Argentina y Uruguay, y la visión de la misma desde el lado de Einstein. En ella se utilizó por primera vez el diario de Einstein sobre Sudamérica. Una versión ampliada de esta monografía aparecerá próximamente en forma de libro.

Ortiz Eduardo L. y Otero, Mario H., "Removiendo el ambiente: La visita de Einstein al Uruguay en 1925", *Mathesis,* Méjico, Nº 2, 1, 2001, pp. 1-35

BIBLIOGRAFÍA RECOMENDADA SOBRE EINSTEIN

Frank Philipp, *Einstein, his life and times*, Jonathan Cape, Londres, 1948. Es una historia seria con un trasfondo filosófico.

Infeld Leopoldo, *Einstein, su obra y su influencia en nuestro mundo*, Lautaro, Buenos Aires, 1961. Trabajo de un colaborador científico de Einstein, que compartió con él diversas campañas políticas.

Lanczos Cornelius, *The Einstein Decade (1905-1915)*, Academic Press, New York, 1965. Es un análisis de los trabajos de Einstein publicados en la *década maravillosa*, preparado por un alumno y colaborador suyo.

Abraham Pais, *"Subtle is the Lord...," The science and life of Albert Einstein*, Oxford University Press, 1982. Este libro es de un físico teórico con un conocimiento profundo de Einstein y de su obra.

Schulmann Robert y Renn Jürgen, *Albert Einstein/Mileva Maric: The love letters*, Princeton University Press, 1992. Este texto aborda aspectos personales de la vida del científico, como por ejemplo la correspondencia con Mileva Maric, su primera esposa.

Podgorny Irina, "La inmigración de científicos alemanes", en *Todo es Historia* Nº 413, diciembre de 2001. Este artículo se ocupa del aporte de los científicos alemanes al desarrollo de la ciencia en la Argentina. Algunos de los nombres que se retratan estaban vinculados con Einstein.

EL DIARIO DE EINSTEIN

Hasta hace muy pocos años, la consulta de las libretas del *Diario de Einstein* era difícil y estaba rodeada de complejas formalidades y compromisos. Hoy, como consecuencia de una actitud más realista de los ejecutores de su herencia, ha sido posible un amplio proyecto de publicación de las obras de Einstein (*The Collected Papers of Albert Einstein*), conocido como el *Einstein Papers Project*. El material del diario referente a Sudamérica es accesible en Internet en la dirección: www.alberteinstein.info. En este trabajo se han reproducido textos tomados de esa fuente y transcripciones de Ortiz (1995).